\documentclass[prb, showpacs]{revtex4}

\begin{document}

\title{On the stability of vortex-plane solitons: The solution of the
problem of Josephson-vortex structure in layered superconductors and stacked
junctions }
\author{ Sergey V. Kuplevakhsky}
\affiliation{Institute for Low Temperature Physics and Engineering, \\
61103 Kharkov, UKRAINE\\
and Department of Physics, Kharkov National University, \\
61077 Kharkov, UKRAINE}

\date{\today}

\begin{abstract}
By determining the type of all stationary points of the Gibbs free energy
functional for layered superconductors in parallel magnetic fields, we
establish the classification of all solutions to coupled static sine-Gordon
equations for the phase differences with respect to their stability. We
prove that the only minimizers of the free energy are the Meissner solution
(the ''vacuum'' state) and soliton vortex-plane solutions [S. V.
Kuplevakhsky, Phys. Rev. B \textbf{60}, 7496 (1999); \textit{ibid.} \textbf{%
63}, 054508 (2001); cond-mat/0202293]. They are the actual equilibrium field
configurations. We present a topological classification of these solutions.
In contrast, previously proposed non-soliton configurations (''isolated
fluxons'', ''triangular Josephson-vortex lattices'', etc.) are absolutely
unstable and unobservable: They are nothing but saddle points of the Gibbs
free-energy functional and are not even stationary points of the Helmholtz
free-energy functional (obtained from the former by a Legendre
transformation). (Physically, non-soliton configurations violate
conservation laws for the current and the flux.) The obtained results allow
us to explain dynamic stability of vortex planes, noticed in numerical
simulations, and to provide a unified interpretation of the available
experimental data. We hope that the paper will stimulate interest in the
subject of specialists in different fields of physics and in applied
mathematics.
\end{abstract}


\maketitle


\section{Introduction}

In this paper, we present the solution of the problem of equilibrium vortex
structure in layered superconductors and stacked Josephson junctions in the
presence of a parallel, static, homogeneous external magnetic field $H$ and
provide a unified interpretation of the available experimental data. Our
approach consists in a rigorous mathematical analysis of the stability of
all types of flux configurations, proposed in the literature, by means of
exact variational methods for microscopic free-energy functionals.\cite
{K99,K01,K1}

Within the framework of these methods, we have previously obtained a
complete classification of all possible static soliton solutions to coupled
sine-Gordon (SG) equations for phase differences both in infinite [$N=\infty 
$, $N$ is the number of superconducting(S) layers] layered superconductors%
\cite{K99,K01} and finite ($N<\infty $) Josephson-junction stacks\cite{K1}
for $H>0$. Based on the fundamental argument of soliton physics\cite
{BP75,Bo78,R82,DEGM82,GM86,S89} that topological solitons in nonlinear field
theories are minimizers of the energy functionals (free-energy functionals
in our case),\cite{r2} we have identified these solutions with equilibrium
Josephson-vortex configurations. Their magnetic field has symmetry typical
of plane defects, hence the term ''vortex planes''. Physically, a vortex
plane can be regarded as a bound state of interlayer vortices (one vortex
per each insulating layer in the plane). In contrast to a deep-rooted belief%
\cite{B73} in an ''analogy'' with Abrikosov vortices in continuum type-II
superconductors, the SG equations for $H>0$ do not admit static soliton
solutions that can be identified with an ''isolated Josephson vortex'' or a
''triangular Josephson-vortex lattice''.

Unfortunately, a wide-spread misunderstanding of the fact that equilibrium
Josephson vortices are nothing but static soliton solutions to the SG
equations incurred a misunderstanding of the exact mathematical results of
Refs. \onlinecite{K99,K01,K1}. In the critical comment, Ref. \onlinecite{Kr02}, V. M.
Krasnov and L. N. Bulaevskii ''disprove'' the conclusions of Refs. \onlinecite{K99,K01} by claiming (in contradiction to the fundamentals of soliton
physics) that vortex-plane solitons ''maximize the free energy''.\cite{r1}
According to Ref. \onlinecite{Kr02}, the ''instability'' of vortex planes is
''similar to the instability of the laminar solution\cite{dG} for type-II
superconductors''. [Alternating superconducting and normal layers, envisaged
by the laminar model,\cite{dG} have nothing to do with soliton physics and
do not possess the property of topological stability: see a proof in
subsection V.B of the present paper.] V. M. Krasnov and L. N. Bulaevskii
insist on hypothetical ''isolated fluxons'', allegedly, having ''lower
energy'' for $H>0$ and characterized by a ''much smaller length scale''. No
exact mathematical definition of the ''isolated fluxons'' is given. We have
not found any definition in the original papers by the critics of vortex
planes, either: For example, it is claimed in Refs. \onlinecite{BCG91,Kr01} that
Josephson vortices ''do not exist'' in single Josephson junctions with $W\ll
2\lambda _J$ ($W$ is the junction width, $\lambda _J$ is the Josephson
length). However, an exact, closed-form analytical solution\cite{K1,K2} to
the single static SG equation, appropriate for this case, clearly
demonstrates the existence of phase-difference solitons for arbitrarily
small $W$, provided the external field $H$ is sufficiently high. [As shown
in Refs. \onlinecite{K99,K01}, exactly these solitons account for the well-known
Fraunhofer pattern of the critical Josephson current $I_c\left( H\right) $
for $W\ll 2\lambda _J$.]

Furthermore, the manuscript Ref. \onlinecite{K1}, although submitted twice to
Physical Review B (November 2000, August 2001), is still not published. In
particular, one of the referees disputed the conclusions of Ref. \onlinecite{K1},
because, in his opinion, the soliton boundary conditions\cite
{Bo78,R82,DEGM82,GM86,S89} employed therein ''overdetermined'' the problem
of the classification of equilibrium Josephson-vortex configurations. He
argued that certain numerical simulations for the SG equations had
demonstrated, aside from vortex-plane solitons, the existence of
''single-vortex'' solutions. According to the referee, these solutions had
''lower free energy'' than the vortex planes for given $H$. As in Ref. \onlinecite{Kr02}, no exact definition of such solutions was given.

It should be emphasized that the idea of an ''analogy'' between the
Josephson-vortex structure in layered superconductors and the
Abrikosov-vortex structure in continuum type-II superconductors was not
supported in Ref. \onlinecite{B73} and subsequent publications\cite
{CC90,BC91,BLK92,Ko93} by any serious mathematical arguments.\cite
{F98,K99,K01} Neither was it confirmed by direct experimental observations
of the equilibrium Josephson-vortex structure in artificial stacked
junctions at $H>0$.\cite{N96,Yu98} Unfortunately, most theoretical efforts
were constrained by the idea of an ''analogy'', hence the use of
mathematically ill-formulated methods, such as, e.g., a ''continuum-limit
approximation''.\cite{B73,CC90,BC91,BLK92,Ko93} For instance, the exact SG
equations for the phase differences were not even derived in Refs. \onlinecite{B73,CC90}, concerned with a ''single Josephson vortex'' at $H>0$. (As shown
by Farid,\cite{F98} equations of Ref. \onlinecite{CC90} have no physical
solution.) Since the problem of the stability of the proposed ''vortex
configurations'' (i.e., whether they are actual points of minima of the
free-energy functionals) required the use of rigorous mathematical methods,
it was not even posed in Refs. \onlinecite{B73,CC90,BC91,BLK92,Ko93}.

Concerning numerical simulations for the static SG equations,\cite{Kr00,Kr01}
there is an unjustified tendency to identify any kink-type feature of the
phase difference with a ''Josephson vortex'', without any analysis of its
stability. In contrast to the exact analytical methods of Refs. \onlinecite{K99,K01,K1}, the numerical approach does not provide any means to establish
a full set of necessary and sufficient conditions of the minimum of the
free-energy functionals. Typically,\cite{Kr00} numerical simulations start
with an incorrectly formulated (both mathematically and physically) boundary
value problem that does not meet the criterion of uniqueness.\cite{CH}

To close the issue of the equilibrium Josephson-vortex structure in layered
superconductors and stacked junctions, we determine analytically (by means
of exact methods of the calculus of variations\cite{L62} and soliton physics%
\cite{ZK74,BP75,R82,DEGM82,GM86,S89}) the type of all stationary points of
the exact microscopic Gibbs free-energy functional,\cite{K1} generating the
static SG equations. Our consideration applies to an arbitrary number of
superconducting layers $N$, including the cases $N=2$ (a single junction)
and $N\rightarrow \infty $\ (an infinite layered superconductor). As a
result, we obtain a complete classification of all nontrivial solutions to
the SG equations, considered in the literature (both analytically and
numerically), with respect to their stability. As could be expected from the
general arguments of soliton physics, the only minimizers of the free-energy
functional are the Meissner solution (the ''vacuum'' state) and soliton
vortex-plane solutions.\cite{K99,K01,K1} The latter represent the actual
equilibrium Josephson-vortex configurations for $H>0$. In contrast,
non-soliton configurations (e.g., ''isolated fluxons'', ''triangular
Josephson-vortex lattices'', etc.) are absolutely unstable: They are nothing
but saddle points of the Gibbs free-energy functional and are not even
stationary points of the Helmholtz free-energy functional (obtained from the
former by a Legendre transformation). [Physically, non-soliton solutions
violate conservation laws for the current and the flux, as was first noticed
in the case of an infinite layered superconductor ($N=\infty $) in Refs. \onlinecite{K99,K01}.]

In section II, we clarify a relationship between the correct formulation of
the boundary value problem to the SG equations and a full set of necessary
and sufficient conditions of the minimum of the Gibbs free-energy
functional. The proof of the stability of the Meissner solution and
vortex-plane solitons is given in sections III, IV. In section III, we
establish that the sufficient condition of the minimum of the Gibbs
free-energy functional consists in the vanishing of the surface variation of
the corresponding Helmholtz free-energy functional, which, in turn, yields
conservation laws for the flux and the intralayer current. In subsection
IV.A, we derive the soliton boundary conditions directly from the
conservation law for the flux, which provides the sought proof of the
stability of the Meissner solution and vortex-plane solitons. The main
physical and mathematical properties of these solutions are discussed in
subsection IV.B. In subsection IV.C, we analyze the obtained results from
the point of view of general theory of topological defects\cite
{BP75,Bo78,M79,R82,DEGM82,GM86,S89} and explain dynamic stability of vortex
planes, established in numerical simulations.\cite{SBP93,Kl94,Ma00}

Some important physical and mathematical issues, related to the main results
of the paper, are discussed in section V. In subsection V.A, we present a
rigorous analytical description of unstable solutions (''isolated fluxons'',
''triangular Josephson-vortex lattices'', etc.), proposed in previous
theoretical publications and numerical simulations. In subsection V.B, we
draw a comparison with the Abrikosov vortices in type-II superconductors. In
subsection V.C, we analyze the available experimental data from the point of
view of the stable vortex-plane configurations.

In section VI, we summarize the obtained results, systematize our criticism
of previous approaches and make some concluding remarks. In Appendix A, we
give a list of mathematical formulas, relevant to the main text. In Appendix
B, we establish a relationship to the variational principle of Refs. \onlinecite
{K99,K01} for infinite layered superconductors.

Throughout the paper, we adhere to the dimensionless notation of Ref. \onlinecite{K1}. The geometry of the problem is that of figures 1, 2 in Ref. \onlinecite{K1}:
The superconductor occupies the region $\left[ 0\leq x\leq N-1\right] \times
\left[ -L\leq y\leq L\right] \times \left( -\infty <z<\infty \right) $,
where $2L=W$; the layering axis (the $c$-axis) is $x$; the axis $y$ is along
the layers; the external magnetic field is along the axis $z$: ${\bf H}%
=\left( 0,0,H\geq 0\right) $. The phase difference between two successive
S-layers is denoted as $\phi _n\equiv \varphi _n-\varphi _{n-1}$ ($\phi
_0=\phi _N\equiv 0$), with $n=0,1,\ldots ,N-1$ being the S-layer number. The 
$c$-axis external current is not considered: $I=0$.

\section{The formulation of the problem}

In Ref. \onlinecite{K1}, we have derived the coupled static SG equations for the
phase differences $\phi _n$ by minimizing the exact microscopic Gibbs
free-energy functional\cite{K01} $\Omega \left[ f_n,\phi _n,{\bf A};H\right] 
$ with respect to $f_n$ and ${\bf A\ }$($f_n$ is the reduced modulus of the
order parameter in the $n$-th S-layer, ${\bf A}$ is the vector potential).
In the limit $r\left( T\right) \ll 1$, $H\ll H_{c2}$ [$r\left( T\right) $ is
the parameter of the interlayer coupling, $H_{c2}$ is the upper critical
field], when $f_n=1$, the SG equations appear as solubility conditions for
the Maxwell equations in the gauge 
\begin{equation}
\label{2.1}{\bf A}=\left[ 0,A\left( x,y\right) ,0\right] . 
\end{equation}
The SG equations read:\cite{K1} 
\begin{equation}
\label{2.2}\frac{d^2\phi _n\left( y\right) }{dy^2}=\frac 1{\epsilon
^2}\sum_{m=1}^{N-1}G^{-1}\left( n,m\right) \sin \phi _m\left( y\right)
,\quad n=1,\ldots ,N-1; 
\end{equation}
where $G^{-1}\left( n,m\right) $ is a Jacobian matrix with elements $%
G^{-1}\left( n,n\right) =2+\epsilon ^2$ ($n=1,\ldots ,N-1$), $G^{-1}\left(
n+1,n\right) =G^{-1}\left( n,n+1\right) =-1$ ($n=1,\ldots ,N-2$), and $%
G^{-1}\left( n,m\right) =0$ for $\left| n-m\right| >1$.

The requirement that the local field $H_n\left( y\right) $ ($n-1\leq x<n$, $%
n=1,\ldots ,N-1$) be equal to the applied one at $y=\pm L$ has led to the
conditions 
\begin{equation}
\label{2.3}\frac{d\phi _n}{dy}\left( -L\right) =\frac{d\phi _n}{dy}\left(
L\right) ,\quad n=1,\ldots ,N-1; 
\end{equation}
\begin{equation}
\label{2.4}\frac{d\phi _n}{dy}\left( -L\right) =\frac{d\phi _{n+1}}{dy}%
\left( -L\right) \equiv \frac{d\phi }{dy}\left( -L\right) \geq 0,\quad
n=1,\ldots ,N-2. 
\end{equation}
[The condition $\frac{d\phi }{dy}\left( -L\right) \geq 0$ merely reflects
the fact that the local field is parallel to the applied one $H\geq 0$. ]
From the requirement that the local field be equal to the applied one at $%
x=0 $, $x=N-1$, we have obtained 
\begin{equation}
\label{2.5}\phi _n\left( y\right) =\phi _{N-n}\left( y\right) ,\quad
n=1,\ldots ,N-1. 
\end{equation}
Equations (\ref{2.2}) and boundary conditions (\ref{2.3}) are satisfied by
functions of the type 
\begin{equation}
\label{2.6}\phi _n\left( y\right) =-\phi _n\left( -y\right) +2\pi Z_n, 
\end{equation}
where the constants $Z_n$ can be arbitrarily chosen from the set $0,\pm
1,\pm 2,\ldots $ As is pointed out in Ref. \onlinecite{K1}, the fixation of the
constants $Z_n$ requires imposition of boundary conditions on $\phi _n$ at $%
y=\pm L$. Based on general arguments of soliton physics that soliton
solutions are minimizers of corresponding energy functionals, we have
imposed in Ref. \onlinecite{K1} standard soliton boundary conditions
on $\phi _n$.

As it has turned out, the only possible solutions, compatible with the
requirement (\ref{2.4}), are the Meissner solution and the soliton
vortex-plane solutions, for which (\ref{2.5}) is satisfied automatically and 
\begin{equation}
\label{2.7}Z_n=Z_{n+1}\equiv N_v,\quad n=1,\ldots ,N-2;\quad
N_v=0,1,2,\ldots 
\end{equation}
in (\ref{2.6}), with $N_v=0$ representing the topologically trivial Meissner
solution. Given that\cite{K1} 
\begin{equation}
\label{2.8}H_n\left( y\right) =H\left[ G\left( n,1\right) +G\left(
n,N-1\right) \right] +\frac{\epsilon ^2}2\sum_{m=1}^{N-1}G\left( n,m\right) 
\frac{d\phi _m\left( y\right) }{dy}, 
\end{equation}
the field $H=\bar H$, corresponding to a concrete configuration $\left\{
\bar \phi _n\right\} $ with $N_v=\bar N_v$, is determined by 
\begin{equation}
\label{2.11}\frac{d\bar \phi }{dy}\left( -L\right) =2\bar H. 
\end{equation}
Note that the first, phase-independent, term in (\ref{2.8}) is a
contribution of the field penetrating through the boundaries $x=0$, $x=N-1$,
and the second term is a contribution of the field penetrating through the
boundaries $y=\pm L$. The matrix $G\left( n,m\right) $is the inverse of $%
G^{-1}\left( n,m\right) $: for its properties, see Appendix A.

In contrast to the above exact variational method, numerical simulations\cite
{Kr00,Kr01} for (\ref{2.2}) start with the imposition of the boundary
conditions 
\begin{equation}
\label{2.9}\frac{d\phi _n}{dy}\left( \pm L\right) =2H, 
\end{equation}
without any regard to sufficient conditions of the minimum of the Gibbs
free-energy functional. Such an approach is based on an erroneous belief
that all solutions to (\ref{2.2}), (\ref{2.9}) minimize the free-energy
functional. (Manifestations of this belief are the naive ''energy
arguments'' of Ref. \onlinecite{Kr02}, appealing to ''a difference in the length
scales'', and calculations by means of combinatorics\cite{Kr00,Kr01} of the
''number of quasi-equilibrium fluxon modes''.) However, conditions (\ref{2.9}%
) do not specify any boundary value problem for (\ref{2.2}): By virtue of
the symmetry relations (\ref{2.6}), the imposition of the boundary condition
on $\frac{d\phi _n}{dy}$ at $y=-L$ automatically ensures the fulfillment of
the same boundary condition at $y=L$, whereas the constants $Z_n$ remain
undetermined. Thus, the ''boundary value problem'' (\ref{2.9}) does not
satisfy the criterion of uniqueness,\cite{CH} which is a sign of the
presence of unphysical (i.e., unobservable) solutions.

The existence of redundant solutions to (\ref{2.2}), (\ref{2.9}) is already
clear for physical reasons: This ''boundary value problem'' does not take
any account of the necessity to ensure the continuity of the local field at
the boundaries $x=0$, $x=N-1$. To understand at a rigorous mathematical
level where the unphysical solutions come from, we have to consider all
stationary points of the generating Gibbs free-energy functional, rewritten
via $\phi _n$ and $\frac{d\phi _n}{dy}$.\cite{K1} In this way, we will
derive a full set of necessary and sufficient conditions of the minimum
directly from the variational principle, obtain an independent proof of the
fact that the Meissner solution and the soliton vortex-plane solutions\cite
{K99,K01,K1} are the unique minimizers of the problem and establish the
character of the instability of unphysical non-soliton solutions. In our
consideration, we will employ the first integral of (\ref{2.2}) that, taking
account of (\ref{2.9}), has the form\cite{K1}%
$$
\sum_{n=1}^{N-1}\cos \phi _n\left( y\right) +\frac{\epsilon ^2}%
2\sum_{n=1}^{N-1}\sum_{m=1}^{N-1}G\left( n,m\right) \frac{d\phi _n\left(
y\right) }{dy}\frac{d\phi _m\left( y\right) }{dy} 
$$
\begin{equation}
\label{2.10}=\frac{2H^2}{H_s^2}\left( N-1\right) +\sum_{n=1}^{N-1}\cos \phi
_n\left( L\right) , 
\end{equation}
where $H_s$ is the superheating (penetration) field of a semiinfinite ($%
0\leq y<+\infty $) Josephson-junction stack, given by Eq. (\ref{a7}).

\section{The necessary and sufficient conditions of the minimum of the Gibbs
free-energy functional}

The generating Gibbs free-energy functional for the SG equations (\ref{2.2})
has the form\cite{K1} 
\begin{equation}
\label{3.1}\Omega \left[ \phi _n,\frac{d\phi _n}{dy};H\right] =F\left[ \phi
_n,\frac{d\phi _n}{dy};H\right] -4Hr\left( T\right) \sum_{n=1}^{N-1}\Phi _n 
\frac{\phi _n\left( L\right) -\phi _n\left( -L\right) }{2\pi }, 
\end{equation}
$$
F\left[ \phi _n,\frac{d\phi _n}{dy};H\right] 
$$
$$
=r\left( T\right) \left[ \frac{2H^2}{H_s^2}W\left( N-1\right) +\frac{%
\epsilon ^2}2\sum_{n=1}^{N-1}\sum_{m=1}^{N-1}G\left( n,m\right)
\int\limits_{-L}^Ldy\frac{d\phi _n\left( y\right) }{dy}\frac{d\phi _m\left(
y\right) }{dy}\right. 
$$
\begin{equation}
\label{3.2}\left. +\sum_{n=1}^{N-1}\int\limits_{-L}^Ldy\left[ 1-\cos \phi
_n\left( y\right) \right] \right] , 
\end{equation}
\begin{equation}
\label{3.3}\Phi _n=\pi \left[ 1-G\left( n,1\right) -G\left( n,N-1\right)
\right] . 
\end{equation}
Note that the functional (\ref{3.1}) is measured from the condensation
energy $\Omega _0=-\frac{NW}2$. Moreover, the energy of the external field
in the absence of the sample is subtracted.\cite{LL}

Our task is to establish the necessary and sufficient conditions of the
minimum of (\ref{3.1}):%
$$
\Delta \Omega \left[ \phi _n,\frac{d\phi _n}{dy};H\right] 
$$
\begin{equation}
\label{3.4}=\Omega \left[ \phi _n+\delta \phi _n,\frac{d\phi _n}{dy}+\frac{%
d\delta \phi _n}{dy};H\right] -\Omega \left[ \phi _n,\frac{d\phi _n}{dy}%
;H\right] \geq 0 
\end{equation}
First, we observe that, in contrast to (\ref{3.1}), the functional (\ref{3.2}%
) is positive definite 
\begin{equation}
\label{3.5}F\left[ \phi _n,\frac{d\phi _n}{dy};H\right] \geq 0, 
\end{equation}
since the matrix $G\left( n,m\right) $ is positive definite: see Appendix A.
The absolute minimum of (\ref{3.2}) is achieved for $H=0$, $\phi _n\equiv 0$
($n=1,2,\ldots ,N-1$). Hence, the functional (\ref{3.2}) necessarily has
minima for any $H\geq 0$. The first variations of (\ref{3.1}), (\ref{3.2})
are 
\begin{equation}
\label{3.6}\delta \Omega \left[ \phi _n,\frac{d\phi _n}{dy};H\right] =\delta
F\left[ \phi _n,\frac{d\phi _n}{dy};H\right] -4Hr\left( T\right)
\sum_{n=1}^{N-1}\Phi _n\frac{\delta \phi _n\left( L\right) -\delta \phi
_n\left( -L\right) }{2\pi }, 
\end{equation}
$$
\delta F\left[ \phi _n,\frac{d\phi _n}{dy};H\right] 
$$
$$
=r\left( T\right) \sum_{n=1}^{N-1}\int\limits_{-L}^Ldy\left[ \sin \phi
_n\left( y\right) -\epsilon ^2\sum_{m=1}^{N-1}G\left( n,m\right) \frac{%
d^2\phi _m\left( y\right) }{dy^2}\right] \delta \phi _n\left( y\right) 
$$
\begin{equation}
\label{3.7}+2\frac{d\phi }{dy}\left( -L\right) r\left( T\right)
\sum_{n=1}^{N-1}\Phi _n\frac{\delta \phi _n\left( L\right) -\delta \phi
_n\left( -L\right) }{2\pi }. 
\end{equation}
The requirement that the volume variation in (\ref{3.7}) vanish yields the
SG equations (\ref{2.2}), as expected. Of special interest to us are surface
variations, i.e., the last terms in (\ref{3.6}), (\ref{3.7}): The
requirement that these variations vanish determines boundary conditions to (%
\ref{2.1}). (For a very clear discussion of the relationship between the
surface variation and boundary conditions, see Ref. \onlinecite{L62}, section
II.15.) In the derivation of the surface variation in (\ref{3.7}), we have
used conditions (\ref{2.3}), (\ref{2.4}) that the local field be continuous
at the boundaries $y=\pm L$. The requirement of the continuity of the local
field at the boundaries $x=0$, $x=N-1$ has not been so far employed. [Recall
our remark in section II that the disregard of this requirement is the
reason for unphysical solutions to (\ref{2.1}), (\ref{2.9}).]

If we simply enforce the conditions (\ref{2.9}), the surface variations in (%
\ref{3.6}) and (\ref{3.7}) cancel out: Thus, all solutions to (\ref{2.2}), (%
\ref{2.9}) are stationary points of the Gibbs free-energy functional (\ref
{3.1}). However, under (\ref{2.9}), the surface variation in (\ref{3.7})
does not vanish. Therefore, not all solutions to (\ref{2.2}), (\ref{2.9})
are stationary points of (\ref{3.2}). We have to examine conditions of the
stationarity of (\ref{3.2}) in more detail.

The requirement that the local field is fixed at the boundaries $x=0$, $%
x=N-1 $ is equivalent to the requirement that the vector potential ${\bf A}$
is fixed at $x=0$, $x=N-1$. Consider now the total flux $\Phi $. In the
gauge (\ref{2.1}), we have 
\begin{equation}
\label{3.8}\Phi =\int\limits_{-L}^Ldy\left[ A\left( N-1,y\right) -A\left(
0,y\right) \right] . 
\end{equation}
On the other hand, 
\begin{equation}
\label{3.9}\Phi =\sum_{n=1}^{N-1}\int\limits_{-L}^LdyH_n\left( y\right)
=HW\left( N-1\right) \frac{H_s^2-1}{H_s^2}+\sum_{n=1}^{N-1}\Phi _n\frac{\phi
_n\left( L\right) -\phi _n\left( -L\right) }{2\pi }, 
\end{equation}
where the first term is the flux penetrating through the boundaries $x=0$, $%
x=N-1$, and the second term is the flux penetrating through the boundaries $%
y=\pm L$. (We will call it the ''Josephson flux'', $\Phi _J$.) Given that 
\begin{equation}
\label{3.10}\delta A\left( 0,y\right) =\delta A\left( N-1,y\right) =0, 
\end{equation}
we have 
\begin{equation}
\label{3.11}\delta \Phi =\delta \Phi _J=\sum_{n=1}^{N-1}\Phi _n\frac 1{2\pi
}\int\limits_{-L}^Ldy\frac{d\delta \phi _n\left( y\right) }{dy}%
=\sum_{n=1}^{N-1}\Phi _n\frac{\delta \phi _n\left( L\right) -\delta \phi
_n\left( -L\right) }{2\pi }=0. 
\end{equation}
Thus, the continuity of the field at $x=0$, $x=N-1$ imposes a constraint on
the variations: 
\begin{equation}
\label{3.12}\Phi _J=\sum_{n=1}^{N-1}\Phi _n\frac 1{2\pi
}\int\limits_{-L}^Ldy \frac{d\phi _n\left( y\right) }{dy}=const\geq 0. 
\end{equation}
The result (\ref{3.12}) is exactly what had to be expected:\cite{LL} By
virtue of the Meissner effect, the flux $\Phi _J$ (and, of course, $\Phi $)
is stable against any small perturbations, represented by variations $\phi
_n\left( y\right) \rightarrow \phi _n\left( y\right) +\delta \phi _n\left(
y\right) $. Equivalent forms of (\ref{3.11}), (\ref{3.12}) are 
\begin{equation}
\label{3.13}\delta \phi _n\left( L\right) =\delta \phi _n\left( -L\right) , 
\end{equation}
\begin{equation}
\label{3.14}\phi _n\left( L\right) -\phi _n\left( -L\right) =c_n=const\geq 0. 
\end{equation}

Note that the existence of conserved physical quantities of the type of $%
\Phi _J$ is a precursor to the existence of soliton solutions in nonlinear
field theories.\cite{ZK74,BP75,Bo78,R82,DEGM82,GM86,S89} In Appendix B, we
establish a relationship between the conservation of $\Phi _J$ and the
conservation of the intralayer current, which, in turn, establishes a
relationship to the variational principle for infinite ($N=\infty $) layered
superconductors.\cite{K99,K01}

What will be shown now is that all the stationary points of (\ref{3.2}) are
the unique minimizers of both (\ref{3.1}) and (\ref{3.2}). First, we notice
that high-order variations of (\ref{3.1}) and (\ref{3.2}) coincide: $\delta
^nF=\delta ^n\Omega $, $n\geq 2$. Thus, all the minimizers of (\ref{3.2})
are minimizers of (\ref{3.1}). On the other hand, the minimizers of (\ref
{3.1}) obeying (\ref{3.12}) are minimizers of (\ref{3.2}): From the
condition of the minimum (\ref{3.4}), we get%
$$
F\left[ \phi _n+\delta \phi _n,\frac{d\phi _n}{dy}+\frac{d\delta \phi _n}{dy}%
;H\right] -F\left[ \phi _n,\frac{d\phi _n}{dy};H\right] -\delta \Phi _J 
$$
\begin{equation}
\label{3.15}=F\left[ \phi _n+\delta \phi _n,\frac{d\phi _n}{dy}+\frac{%
d\delta \phi _n}{dy};H\right] -F\left[ \phi _n,\frac{d\phi _n}{dy};H\right]
\geq 0. 
\end{equation}
[Physically, this fact means the equivalence of the description in terms of
the Gibbs free energy and the Helmholtz free energy: Because of (\ref{3.12}%
), the functional $F\left[ \phi _n,\frac{d\phi _n}{dy};\Phi _J\right] \equiv
F\left[ \phi _n,\frac{d\phi _n}{dy};0\right] $ can be regarded as the
Helmholtz free-energy functional.] Using the standard technique,\cite{ZK74}
it is straightforward to prove that all stationary points of (\ref{3.1}),
obeying (\ref{3.12}), are minimizers of (\ref{3.1}) [and, thus, of (\ref{3.2}%
)]. Indeed, let $\left\{ \bar \phi _n\right\} $ be the stationary point for $%
\Phi _J=\bar \Phi _J$ and corresponding $H=\bar H$. In the vicinity of $%
\left\{ \bar \phi _n\right\} $, i.e., for $\phi _n=\bar \phi _n+\delta \bar
\phi _n$, we have the following estimate:%
$$
\Omega \left[ \phi _n,\frac{d\phi _n}{dy};\bar H\right] 
$$
$$
\geq r\left( T\right) \left[ \frac{\epsilon ^2}2\sum_{n=1}^{N-1}%
\sum_{m=1}^{N-1}G\left( n,m\right) \int\limits_{-L}^Ldy\frac{d\phi _n\left(
y\right) }{dy}\frac{d\phi _m\left( y\right) }{dy}-4\bar
H\sum_{n=1}^{N-1}\Phi _n\frac 1{2\pi }\int\limits_{-L}^Ldy\frac{d\phi
_n\left( y\right) }{dy}\right] 
$$
\begin{equation}
\label{3.16}\geq -4\bar Hr\left( T\right) \bar \Phi _J. 
\end{equation}
Inequalities (\ref{3.16}) show that $\Omega $ has a lower bound in the
vicinity of any stationary point $\left\{ \bar \phi _n\right\} $, obeying (%
\ref{3.12}); hence, $\left\{ \bar \phi _n\right\} $ is a minimizer of $%
\Omega $ and $F$ , Q.E.D.

To strengthen (\ref{3.16}), we minimize the right-hand side of the first
inequality with respect to $\frac{d\phi _n}{dy}$, obtaining 
\begin{equation}
\label{3.17}\epsilon ^2\sum_{m=1}^{N-1}G\left( n,m\right) \left[ \frac{d\phi
_m}{dy}\right] _{\min }=\frac{2\bar H}\pi \Phi _n, 
\end{equation}
\begin{equation}
\label{3.17.1}\Omega \left[ \phi _n,\frac{d\phi _n}{dy};\bar H\right] \geq
-2\bar Hr\left( T\right) \bar \Phi _J. 
\end{equation}
Taking into account that $\left\{ \bar \phi _n\right\} $ is a solution of (%
\ref{2.2}), making use of (\ref{2.10}) and (\ref{3.17}), we get: 
$$
\Omega \left[ \phi _n,\frac{d\phi _n}{dy};\bar H\right] \geq \Omega \left[
\bar \phi _n,\frac{d\bar \phi _n}{dy};\bar H\right] 
$$
$$
=r\left( T\right) \left[ W\sum_{n=1}^{N-1}\left[ 1-\cos \bar \phi _n\left(
L\right) \right] +\epsilon ^2\sum_{n=1}^{N-1}\sum_{m=1}^{N-1}G\left(
n,m\right) \int\limits_{-L}^Ldy\frac{d\bar \phi _n\left( y\right) }{dy}\frac{%
d\bar \phi _m\left( y\right) }{dy}-4\bar H\bar \Phi _J\right] 
$$
\begin{equation}
\label{3.18}\geq r\left( T\right) W\sum_{n=1}^{N-1}\left[ 1-\cos \bar \phi
_n\left( L\right) \right] \geq 0. 
\end{equation}
The inequality $\Omega \left( \bar \phi _n,\frac{d\bar \phi _n}{dy};\bar
H\right) \geq 0$ is a manifestation of the Meissner effect and had to be
expected from general thermodynamic arguments.\cite{LL} Relations (\ref{3.18}%
) immediately yield%
$$
F\left[ \phi _n,\frac{d\phi _n}{dy};\bar H\right] \geq F\left[ \bar \phi _n, 
\frac{d\bar \phi _n}{dy};\bar H\right] 
$$
\begin{equation}
\label{3.19}\geq r\left( T\right) W\sum_{n=1}^{N-1}\left[ 1-\cos \bar \phi
_n\left( L\right) \right] +4\bar Hr\left( T\right) \bar \Phi _J\geq 4\bar
Hr\left( T\right) \bar \Phi _J. 
\end{equation}
We want to emphasize that inequalities of the type (\ref{3.17.1}), (\ref
{3.19}) are typical of soliton physics: They are used to establish the
existence and stability of soliton solutions.\cite
{ZK74,BP75,R82,DEGM82,GM86,S89}

Note an alternative interpretation of the variational principle 
\begin{equation}
\label{3.20}\delta \Omega \left[ \phi _n,\frac{d\phi _n}{dy};H\right]
=\delta F\left[ \phi _n,\frac{d\phi _n}{dy};0\right] -4Hr\left( T\right)
\sum_{n=1}^{N-1}\Phi _n\frac 1{2\pi }\delta \int\limits_{-L}^Ldy\frac{d\phi
_n\left( y\right) }{dy}=0. 
\end{equation}
The field $H$ in (\ref{3.20}) can be considered as a Lagrange multiplier,
implying that variation can be performed without any restrictions on $\delta
\phi _n\left( L\right) $, $\delta \phi _n\left( -L\right) $. In this case,
the requirement that the surface variation vanish yields conditions (\ref
{2.3}), (\ref{2.4}). Boundary conditions on $\phi _n$ are uniquely
determined by (\ref{3.12}): see the next section. The value $H=\bar H$ for a
concrete minimizer $\left\{ \bar \phi _n,\bar \Phi _J\right\} $ should be
found from the condition of thermodynamic equilibrium 
\begin{equation}
\label{3.21}\frac{\partial \Omega \left( \bar \phi _n,\frac{d\bar \phi _n}{dy%
};\bar H\right) }{\partial \bar \Phi _J}=0. 
\end{equation}
Indeed, $\Omega \left( \bar \phi _n,\frac{d\bar \phi _n}{dy};\bar H\right) $
can be written as%
$$
\Omega \left( \bar \phi _n,\frac{d\bar \phi _n}{dy};\bar H\right) =r\left(
T\right) \left[ W\sum_{n=1}^{N-1}\left[ 1-\cos \bar \phi _n\left( L\right)
-\frac 1W\int\limits_{-L}^Ldy\bar \phi _n\left( y\right) \sin \bar \phi
_n\left( y\right) \right] \right. 
$$
\begin{equation}
\label{3.22}\left. +2\left[ \frac{d\bar \phi }{dy}\left( L\right) -2\bar
H\right] \bar \Phi _J\right] , 
\end{equation}
which by virtue of (\ref{3.21}) immediately yields (\ref{2.11}).

In summary, we have proved the following: The SG equations (\ref{2.1}) and
relations (\ref{2.9}) ensure only the stationarity of the Gibbs free-energy
functional (\ref{3.1}). The necessary and sufficient conditions of the
minimum of both (\ref{3.1}) and (\ref{3.2}) (which is the Helmholtz
free-energy functional for $H=0$) are given by (\ref{2.3}), (\ref{2.4}) and
the constraint (\ref{3.12}). Solutions to (\ref{2.2}), (\ref{2.9}) that do
not obey this constraint are absolutely unstable. The character of this
instability can be easily established. Indeed, such solutions are not even
stationary points of the Helmholtz free-energy functional, therefore $\delta
^2\Omega =\delta ^2F$ need not have definite sign. Moreover, the functional (%
\ref{3.1}) is unbounded in the vicinity of these solutions. Thus, they are
nothing but saddle points of (\ref{3.1}).

\section{The proof of the stability of the Meissner solution and
vortex-plane solitons}

\subsection{Boundary conditions on $\phi _n$}

Our aim now is to establish boundary conditions on $\phi _n$ directly from
the constraint (\ref{3.12}). Given that the SG equations (\ref{2.2}),
boundary conditions (\ref{2.3}), (\ref{2.4}) and the constraint (\ref{3.12})
represent a full set of necessary and sufficient conditions of the minimum
of the Gibbs and Helmholtz free-energy functionals, we will obtain, in this
manner, the sought proof of the stability of the Meissner solution and
soliton vortex-plane solutions.

First, we observe that since a minimizer of (\ref{3.1}), (\ref{3.2}) must
ensure the vanishing of both the surface and volume variations in (\ref{3.6}%
) and (\ref{3.7}), it should necessarily belong to the class of functions
that satisfy (\ref{2.2}), (\ref{2.3}), (\ref{2.4}) and the symmetry
relations (\ref{2.6}). Thus, the variation of the surface terms in (\ref{3.6}%
) and (\ref{3.7}) is performed with respect to trial functions that take
only discrete values at $y=0$: 
\begin{equation}
\label{4.1}\phi _n\left( 0\right) =\pi Z_n, 
\end{equation}
where $Z_n$ can be arbitrarily chosen from the set $0,\pm 1,\pm 2,\ldots $
.These functions can be subdivided into classes parameterized by an $\left(
N-1\right) $-dimensional ''vector'' 
\begin{equation}
\label{4.2}{\bf Q}=\left( Z_1,Z_2,\ldots ,Z_{N-1}\right) . 
\end{equation}

In view of (\ref{4.1}), the requirement of the continuity of variations can
be met if and only if 
\begin{equation}
\label{4.3}\delta \phi _n\left( 0\right) =0, 
\end{equation}
which means that all the minima of (\ref{3.1}), (\ref{3.2}) are
parameterized by the vector ${\bf Q}$, and the variation of the surface
terms in (\ref{3.6}) and (\ref{3.7}) is performed with respect to trial
functions that belong to a certain class (\ref{4.2}). Moreover, the symmetry
relations (\ref{2.6}) imply $\delta \phi _n\left( L\right) =-\delta \phi
_n\left( -L\right) $. Combined with (\ref{3.3}), this yields 
\begin{equation}
\label{4.4}\delta \phi _n\left( L\right) =\delta \phi _n\left( -L\right) =0. 
\end{equation}

Now we combine (\ref{3.4}) with (\ref{2.6}) to obtain 
\begin{equation}
\label{4.5}\phi _n\left( L\right) -\pi Z_n=\frac{c_n}2\geq 0, 
\end{equation}
\begin{equation}
\label{4.6}-\phi _n\left( -L\right) +\pi Z_n=\frac{c_n}2\geq 0. 
\end{equation}
Since the inequalities in (\ref{4.5}), (\ref{4.6}) should hold for any $Z_n$%
, including $Z_n=0$, we get 
$$
\phi _n\left( -L\right) \leq 0,\quad \phi _n\left( L\right) \geq 0. 
$$
Moreover, since for any fixed set $\left\{ c_n\right\} $ the set $\left\{
\phi _n\left( \pm L\right) \right\} $ must belong to a certain unique class (%
\ref{4.2}), 
\begin{equation}
\label{4.7}-\pi <\phi _n\left( -L\right) \leq 0, 
\end{equation}
\begin{equation}
\label{4.8}2\pi Z_n\leq c_n<2\pi \left( Z_n+1\right) ,\quad Z_n=\left[ \frac
1{2\pi }\int\limits_{-L}^Ldy\frac{d\phi _n\left( y\right) }{dy}\right]
=0,1,2,\ldots , 
\end{equation}
where $\left[ u\right] $ is the integer part of $u$.

Given (\ref{4.3}), (\ref{4.4}), the boundary conditions (\ref{4.7}) and 
\begin{equation}
\label{4.9}\phi _n\left( 0\right) =\pi Z_n,\quad Z_n=0,1,2,\ldots 
\end{equation}
together with (\ref{2.3}), (\ref{2.4}) determine, in principle, a complete
set of conditions for the minimizer of (\ref{3.1}), (\ref{3.2}). The
solution belonging to a certain class (\ref{4.2}) [with $Z_n$ as in (\ref
{4.9})] first appears when all $\phi _n\left( -L\right) =0$, and $c_n=2\pi
Z_n$ [see the estimates (\ref{3.18}), (\ref{3.19}): under these conditions,
the density of the Josephson energy at the boundary is a minimum]. Thus, we
are confronted with the standard\cite{Bo78,R82,DEGM82,GM86,S89} soliton
boundary value problem 
\begin{equation}
\label{4.10}\phi _n\left( -L\right) =0,\quad \phi _n\left( 0\right) =\pi
Z_n,\quad Z_n=0,1,2,\ldots , 
\end{equation}
plus the boundary conditions (\ref{2.4}). To find out what type of
minimizers can be realized in reality, we have to solve Eqs. (\ref{2.2}).%
\cite{K1,K2} Note that $N-2$ relations (\ref{2.4}) implicitly impose $N-2$
conditions on $N-1\,$ constants $Z_n$. Therefore, as shown in Refs. \onlinecite
{K1,K2}, the only solutions to (\ref{2.2}), (\ref{4.10}), compatible with (%
\ref{2.4}), are those that satisfy (\ref{2.7}), i.e., the Meissner solution
and the soliton vortex-plane solutions: 
\begin{equation}
\label{4.11}{\bf Q}_v=\left( N_v,N_v,\ldots ,N_v\right) ,\quad N_v=\left[
\frac 1{2\pi }\int\limits_{-L}^Ldy\frac{d\phi _n\left( y\right) }{dy}\right]
=0,1,2,\ldots 
\end{equation}

The properties of these solutions will be discussed in more detail in what
follows. Here we want to brief on the results for $H=0$.\cite{K1,K2} For $%
H=0 $, $L<\infty $, there are no soliton solutions at all (including the
vortex planes), and the only stable solution is the trivial Meissner
solution $\phi _1=\phi _2=\ldots \phi _{N-1}\equiv 0$. The situation changes
drastically for $H=0$, $L=\infty $: The imposition of soliton boundary
conditions on $\phi _n\left( \pm \infty \right) $ automatically ensures the
fulfillment of the boundary conditions $\frac{d\phi _n}{dy}\left( \pm \infty
\right) =0$ by virtue of Eqs. (\ref{2.2}) themselves and some elementary
theorems of mathematical analysis. Aside from the vortex-plane solution with 
$N_v=1$ in (\ref{4.11}) (where $L=\infty $), we have a variety of soliton
solutions (\ref{4.2}) with $Z_n$ arbitrarily chosen from the set $0,\pm 1$.
The fact that for $H=0$, $L=\infty $ each $\phi _n$ can ''accommodate'' no
more than one vortex or anivortex is a generalization of the well-known\cite
{Bo78,R82} result for the single static SG equation and can be easily proved
by the use of the first integral (\ref{2.10}) with the right-hand side equal
to $N-1$.

\subsection{The Meissner solution and soliton vortex-plane solutions}

The range of the existence of the solutions, parameterized by (\ref{4.11}),
is determined by the boundary value problem (\ref{4.7}), (\ref{4.9}) (with
all $Z_n=N_v$) and relation (\ref{4.11}):\cite{K1,K2} 
\begin{equation}
\label{4.12}0\leq H<H_0\equiv H_{sL},\quad for\ N_v=0, 
\end{equation}
\begin{equation}
\label{4.13}\sqrt{H_{N_v-1}^2-H_s^2}\leq H<H_{N_v},\quad for\ N_v=1,2,\ldots , 
\end{equation}
where $H_{sL}$ has the meaning of the superheating field of the Meissner
state ($N_v=0$) for $L<\infty $ ($H_{sL}>H_s$ for $L<\infty $, and $%
H_{s\infty }\equiv H_s$). The lower bound in (\ref{4.12}), (\ref{4.13}) is
determined by the exact upper bound in (\ref{4.7}). At $H=\sup H=H_{N_v}$,
when all $\phi _n\left( -L\right) =\inf \phi _n\left( -L\right) =-\pi $,
there is instability of the saddle-point type (see the end of section III).
Note that both the Meissner solution ($N_v=0$) and the vortex-plane
solutions ($N_v\geq 1$) automatically satisfy the symmetry relations (\ref
{2.5}).\cite{K1,K2}

It is instructive to verify the general inequalities (\ref{3.18}), (\ref
{3.19}). Mathematically, it is sufficient to do this only for $H$ equal to
the lower bounds in (\ref{4.12}), (\ref{4.13}): By continuity arguments, the
result will be valid in the whole field range. For the Meissner solution the
verification is trivial. For $N_v\geq 1$, we employ the exact expression 
\begin{equation}
\label{4.14}F\left[ \bar \phi _n,\frac{d\bar \phi _n}{dy};\bar H\right]
=r\left( T\right) \epsilon ^2\sum_{n=1}^{N-1}\sum_{m=1}^{N-1}G\left(
n,m\right) \int\limits_{-L}^Ldy\frac{d\bar \phi _n\left( y\right) }{dy}\frac{%
d\bar \phi _m\left( y\right) }{dy}, 
\end{equation}
where $\bar H\equiv \sqrt{H_{N_v-1}^2-H_s^2}$. As shown in Ref. \onlinecite{K1},
in this case 
\begin{equation}
\label{4.15}\left[ \frac{d\bar \phi _n\left( y\right) }{dy}\right] _{\min }= 
\frac{d\bar \phi _n}{dy}\left( \pm L\right) =2\sqrt{H_{N_v-1}^2-H_s^2},\quad for\ all\ n. 
\end{equation}
Combining Eqs. (\ref{4.14}), (\ref{4.15}), we get exactly the lower bound in
(\ref{3.19}).

As is emphasized in Ref. \onlinecite{K1}, the obtained solutions are valid for any 
$N$, including the cases $N=2$ (a single junction) and $N=3$ (a
double-junction stack). (For $N=2,3$, we have derived in Ref. \onlinecite{K1}
exact, closed-form analytical expressions.) The solutions with $N_v\geq 1$
are pure solitons only at $H=\sqrt{H_{N_v-1}^2-H_s^2}$, when $j_n\left( \pm
L\right) =\sin \phi _n\left( \pm L\right) =0$ [$j_n\left( y\right) $ is the
density of the Josephson current\cite{K1} for $f_n=1$]. In the rest of the
regions (\ref{4.13}), we have solitons ''dressed'' by the Meissner field. In
the case of $N=2$ (the single junction), Owen and Scalapino\cite{OS67}
called these regions the ''$N_v$ to $N_v+1$ vortex mode''. [Because the
principle of superposition does not apply to the nonlinear Eqs. (\ref{2.2}),
the Meissner and the vortex-plane fields cannot be separated from each
other.] It is clear that the vortex-plane solutions for $N>2$ are a direct
generalization of ordinary vortices in single junctions.

Of special interest is the overlap of the regions (\ref{4.12}), (\ref{4.13})
for $N_v=\bar N_v$ and $N_v=\bar N_v+1$. As a result, the obtained solutions
cover the whole field range $0\leq H<\infty $, as they should.
Mathematically, the overlap is related to the fact that the solution with $%
N_v=\bar N_v$ cannot be continuously transformed into the solution with $%
N_v=\bar N_v+1$ by changing $H$, as is always the case for solitons. For the
single junction, the overlap was first established numerically in Ref. \onlinecite{OS67} and discussed qualitatively in Ref. \onlinecite{K70}. The overlap
practically vanishes for $H_{N_v}\gg H_s$ . Given that all $H_{N_v}$
decrease when $W=2L$ increases,\cite{K1} the overlap is stronger for large $%
W $ and can involve several neighboring states. Physically, the actual
equilibrium state is the one that corresponds to the absolute minimum of the
Gibbs free energy for given $H$. The rest of the allowed states are
metastable. In view of the above-mentioned discontinuity, a transition from
the state with $N_v=\bar N_v$ to the state with lower Gibbs free energy $%
N_v=\bar N_v+1$ will necessarily be a phase transition of the first-order
type.\cite{K70,K99,K01} It particular, the lower critical field $H_{c1}$ is
determined from the requirement that the Gibbs free energy of the state $%
N_v=1$ be equal to that of the Meissner state ($N_v=0$) and satisfies the
relation $\sqrt{H_{sL}^2-H_s^2}<H_{c1}<H_{sL}$.

\subsection{Topological considerations and stability in the dynamic regime}

The stability of the Meissner solution and vortex-plane solitons can be
better understood, if we analyze the obtained results from the general point
of view of the stability of topological defects in continuum media.\cite
{BP75,Bo78,M79,R82,DEGM82,GM86,S89} To this end, we consider the density of
the Gibbs free energy (\ref{3.1}) at the boundaries $y=\pm L$.

Because of the general symmetry relations (\ref{2.6}), valid for any
solution to (\ref{2.6}), (\ref{2.9}), the density of the Josephson energy is
equal at $y=-L$ and $y=+L$: 
\begin{equation}
\label{4.16}1-\cos \phi _n\left( -L\right) =1-\cos \phi _n\left( +L\right)
,\quad n-1\leq x<n,\quad n=1,2,\ldots ,N-1. 
\end{equation}
Taking into account (\ref{2.10}), we conclude that also the density of the
total free energy is equal at the boundaries $y=-L$ and $y=+L$ and thus
corresponds to the degenerate equilibrium (''vacuum'') state, unperturbed by
topological defects (solitons). Mathematically, the boundary of the interval 
$-L\leq y\leq L$ can be considered as a $0$-dimensional sphere: $S^0=\left\{
-L,+L\right\} $. Given that configurations $\phi _n$ and $\phi _n+2\pi Z_n$ (%
$Z_n=0,\pm 1,\pm 2,\ldots $) are physically indistinguishable, we can fix
the values $\phi _n\left( -L\right) $ as in (\ref{4.7}) and regard the
functions 
\begin{equation}
\label{4.17}\psi _n\left( +L\right) \equiv \frac{\phi _n\left( +L\right)
+\phi _n\left( -L\right) }{2\pi }=Z_n 
\end{equation}
as continuous maps of the boundary into the additive group of the integers, $%
{\bf Z}$: $S^0\stackrel{\psi _n}{\rightarrow }{\bf Z}$. (${\bf Z}$ is the
group of the degeneracy of the equilibrium state, or the order-parameter
space.) The fact of the existence of topologically nontrivial maps of this
type, realized by soliton solutions, is often expressed in terms of the
''zeroth homotopy group''\cite{BP75,Bo78,M79,R82,DEGM82,GM86,S89} $\pi
_0\left( M\right) $, where the index ''$0$'' stands for the boundary $S^0$
and $M$ is the order-parameter space: 
\begin{equation}
\label{4.18}\pi _0\left( {\bf Z}\right) ={\bf Z}. 
\end{equation}
[Note that $\pi _0\left( M\right) $ is merely the set of disconnected
components of the space $M$.] Because of the boundary conditions (\ref{2.4}%
), all $\psi _n$ at $H>0$ realize the same mapping : $Z_1=Z_2=\ldots
=Z_{N-1}\equiv Z$. The external field $H>0$ breaks the symmetry $\phi
_n\rightarrow -\phi _n$ [see the second term in (\ref{3.1})]. Therefore,
only the values $Z\equiv N_v=0,1,2,\ldots $ are allowed, with $N_v=0$ being
the ''vacuum'', Meissner state. In this way, we arrive at the natural
topological classification (\ref{4.11}) of the minimizers of (\ref{3.1}).
Owing to the continuity conditions (\ref{4.2}), (\ref{4.3}), variation in (%
\ref{3.6}), (\ref{3.7}) is allowed only with respect to trial functions $%
\left\{ \phi _n\right\} $ that have the same end points $\phi _n\left( \pm
L\right) $ and the middle point $\phi _n\left( 0\right) $ as the minimizer $%
\left\{ \bar \phi _n\right\} $, i.e., $\left\{ \phi _n\right\} $ are
homotopic to $\left\{ \bar \phi _n\right\} $ and belong to the same class (%
\ref{4.11}): hence the stability of $\left\{ \bar \phi _n\right\} $ against
continuous perturbations.

Numerical simulations for time-dependent coupled SG equations have revealed
exceptional stability of vortex planes in the dynamic regime as well: see
figure 7 in Ref. \onlinecite{SBP93}, figure 8 in Ref. \onlinecite{Kl94}, and figure 3 in
Ref. \onlinecite{Ma00}. (The authors of Refs. \onlinecite{SBP93,Kl94,Ma00} employ the
terms ''coherent'', ''in-phase'' or ''phase-locked modes'' instead of our
term ''vortex planes'' that we prefer for physical reasons.) Although a
detailed analysis of the dynamics of vortex planes is beyond the scope of
this paper and will be done elsewhere, the results of Refs. \onlinecite
{SBP93,Kl94,Ma00} can be explained already at this stage.

In the absence of dissipation, the dynamic SG equations, describing an
evolution of the system in the time interval $t_i\leq t\leq t_f$, can be
derived from a corresponding Lagrangian by use of a variational principle.
The requirement that the surface variation vanish on the whole perimeter of
the space-time boundary leads to the conditions $\delta \phi _n\left(
y,t_i\right) =\delta \phi _n\left( y,t_f\right) =0$ and a generalization of
the conservation law for the Josephson flux $\Phi _J$, Eq. (\ref{3.12}): 
\begin{equation}
\label{4.20}\Phi _J=\sum_{n=1}^{N-1}\Phi _n\frac 1{2\pi
}\int\limits_{-L}^Ldy \frac{\partial \phi _n\left( y,t\right) }{\partial y}%
=\sum_{n=1}^{N-1}\Phi _n \frac{\phi _n\left( L,t\right) -\phi _n\left(
-L,t\right) }{2\pi }=const, 
\end{equation}
which means that the differences $\phi _n\left( L,t\right) -\phi _n\left(
-L,t\right) $ do not depend on $t$. Thus, by fixing the boundary conditions $%
\phi _n\left( \pm L,t\right) $ at $t=t_i$ as in subsection IV.A, we fix the
initial value of the flux $\Phi _J=\bar \Phi _J$ that will not change in the
course of the evolution of the system from $t=t_i$ to $t=t_f$. The
topological type of the solution [see (\ref{4.12})] will not change, either: 
\begin{equation}
\label{4.21}N_v=\left[ \frac 1{2\pi }\int\limits_{-L}^Ldy\frac{\partial \phi
_n\left( y,t\right) }{\partial y}\right] =\left[ \frac{\phi _n\left(
L,t\right) -\phi _n\left( -L,t\right) }{2\pi }\right] =const. 
\end{equation}
As usual,\cite{Bo78,R82,DEGM82,GM86,S89} this situation can be formalized in
terms of the conserved topological current 
\begin{equation}
\label{4.22}j_\mu =\sum_{n=1}^{N-1}\frac{\Phi _n}{2\pi }\epsilon _{\mu \tau
}\partial _\gamma \phi _n,\quad \partial _\mu j_\mu =0, 
\end{equation}
where $\mu ,\gamma =0,1$; $\partial _\mu =\left( \partial _t,\partial
_x\right) $; and $\epsilon _{\mu \tau }$ is the antisymmetric symbol on two
indices, $\epsilon _{01}=-\epsilon _{10}=1$; with $\Phi
_J=\int\limits_{-L}^Ldyj_0$ being the topological charge.

As should be clear from these results, time-dependent SG equations alone
cannot describe Josephson-vortex penetration, i.e., an evolution of the
system from the topologically trivial Meissner state, $N_v=0$, to a state
with $N_v\neq 0$. Unfortunately, this important issue has not been realized
in Ref. \onlinecite{Kr02} that claims to have ''demonstrated a dynamic process of
vortex penetration'' by means of numerical simulations for time-dependent SG
equations.

\section{Discussion}

\subsection{Unstable solutions to the SG equations}

As is proved in section III, all non-topological, non-soliton solutions to (%
\ref{2.2}), (\ref{2.9}) that do not meet the requirement of the flux
conservation (\ref{3.12}) are absolutely unstable: They are nothing but
saddle points of the Gibbs free-energy functional (\ref{3.1}), cannot be
assigned any ''free energy'' and are therefore unobservable. Since the
requirement of the continuity of variations in (\ref{3.6}) does not impose
on such solutions any constraints of the type (\ref{4.2}), (\ref{4.3}), they
can be continuously transformed into the stable Meissner solution or a
vortex-plane solution, representing the actual minimum of (\ref{3.1}) at a
given $H$, by a series of infinitesimal deformations of $\phi _n$ without a
violation of the boundary conditions (\ref{2.9}). [A clear illustration of
such a transformation for non-topological defects in a system of planar
spins see in Ref. \onlinecite{M79}, section II.B.]

Analytically, all unstable configurations for $H\geq 0$ can be obtained
using the symmetry relations (\ref{2.6}), as solutions to the boundary value
problem 
\begin{equation}
\label{4.24}\frac{d\phi _n}{dy}\left( -L\right) =2H,\quad \phi _n\left(
0\right) =\pi Z_n,\quad Z_n=0,1,2,\ldots 
\end{equation}
that violate topological boundary conditions, derived in subsection IV.A.
For example, one can set in (\ref{4.24}) $Z_1=Z_2=\ldots =Z_{N-1}\equiv Z$
and increase $H$ beyond the upper bound $H_{N_v}$ of the stability regions (%
\ref{4.12}), (\ref{4.13}) for a given $N_v=Z$. By continuously increasing $H$
beyond the stability region of the Meissner state, unstable configurations with $%
Z=0 $, interpreted as ''Josephson-vortex penetration'', were obtained in
numerical simulations for static SG equations (Ref. \onlinecite{Kl94}, figure 7)
and time-dependent SG equations (Ref. \onlinecite{Kr02}, figure 2). Analogous
instability for $Z\geq 1$ is demonstrated by numerical simulations in the
dynamic regime [$H=$const $>0$, and an increasing transport current $%
I>I_c\left( H\right) $] in Ref. \onlinecite{Ma00}: see region IV in figure 3
therein; region III corresponds to dynamically stable vortex planes.

Unstable solutions appear also when not all $Z_n$ in (\ref{4.24}) are equal
to each other. Thus, an unstable ''single Josephson vortex'' corresponds to
the choice $Z_l=1$, $Z_{n\neq l}=0$. Solutions of this type were obtained in
several numerical simulations.\cite{SBP93,Kr00,Kr01} By way of illustration,
we consider here only the case $H=0$. As is explained at the end of
subsection IV.A, the only stable configuration for $H=0$, $L<\infty $ is the
trivial Meissner state $\phi _1=\phi _2=\ldots =\phi _{N-1}\equiv 0$. Figure
5 in Ref. \onlinecite{SBP93}, and figures 1, 2 in Ref. \onlinecite{Kr01} clearly show
that the solutions presented therein, in reality, are characterized by all $%
Z_n=\left[ \frac 1{2\pi }\int\limits_{-L}^Ldy\frac{d\phi _n\left( y\right) }{%
dy}\right] =0$ and, thus, belong to the class ${\bf Q}_v=\left( 0,0,\ldots
,0\right) $ of the general topological classification (\ref{4.11}). By means
of continuous deformations, they can be transformed into the trivial
Meissner solution.

Other unstable solutions, available in the literature, can be analyzed along
the same lines. In particular, the ''triangular Josephson-vortex lattice
with the period $x_p=1$'', proposed in Ref. \onlinecite{BC91}, corresponds to the
case $Z_{odd}=Z$, $Z_{even}=Z+1$.

\subsection{A comparison with Abrikosov vortices in type-II superconductors}

As the formation of a vortex-plane soliton involves only phase differences
between successive S-layers, it does not affect the topology of the layered
superconductor. In contrast, the appearance of a linear (${\bf R}^1$)
singularity of the order parameter $\Delta \left( {\bf r}\right) =\left|
\Delta \left( {\bf r}\right) \right| \exp \left[ i\varphi \left( {\bf r}%
\right) \right] $ is necessary for the formation of an Abrikosov vortex in
continuum type-II superconductors. Thus, in the presence of a single
Abrikosov vortex, the topology of the continuum type-II superconductor
changes from ${\bf R}^3$ (the three-dimensional Euclidean space) to ${\bf R}%
^3/{\bf R}^1=S^1$. Therefore, the notion of the ''vortex core''\cite{dG} is
inherent (both physically and mathematically) to the Abrikosov vortex and is
meaningless in the case of the vortex plane.

An isolated Abrikosov vortex is itself a stable object, both topologically
and energetically. (The latter can be proved by the use of the same
mathematical methods as those employed in our section III: see, e.g.,
Refs. \onlinecite{DEGM82,GM86,S89}.) Therefore, an equilibrium state
of $N_v$ Abrikosov vortices is determined by comparing the values of the
Ginzburg-Landau free-energy functional for different spatial configurations,
which yields the well-known triangular lattice as the most favorable one.\cite{dG}
In contrast, the notion of the ''Josephson-vortex lattice'' is senseless for layered
superconductors: One can only speak of $N_v$-soliton (vortex-plane) states,
with $N_v=0$ representing the Meissner state, and each vortex plane being
a ''Josephson vortex'' itself.

In the case of extreme type-II superconductors, the linear singularities,
associated with Abrikosov vortices, can be easily incorporated into the
framework of the simple London model.\cite{dG} The resulting equation is a
linear inhomogeneous partial differential equation for the local field.
Owing to linearity, the local field is a superposition of the Meissner and
vortex fields. As is emphasized in subsection IV.B, this is not the case for
layered superconductors because of the nonlinearity of the SG equations (\ref
{2.2}). Unfortunately, this important issue was not understood in some
publications concerned with Josephson-vortex penetration.\cite{BF92}

As we can see, there is no ''analogy'' between the Abrikosov-vortex
structure in continuum type-II superconductors and the Josephson-vortex
structure in layered superconductors in the naive sense.\cite{B73} Instead,
there is a much subtler mathematical analogy: The topological classification
of vortex configurations in type-II superconductors is isomorphic to that in
layered superconductors. A proof is straightforward. For the reasons
explained above, the boundary of a type-II superconductor is, in general,
topologically equivalent to a one-dimensional sphere (a circle) $S^1$. The
order parameter space is $M=U\left( 1\right) $ (the symmetry group of
quantum electrodynamics). Topologically, $U\left( 1\right) =S^1$. Thus,
soliton solutions, in this case, realize nontrivial maps $S^1\rightarrow S^1$%
. All the continuous maps $S^1\rightarrow S^1$ have a group structure of the
fundamental (or first homotopy) group $\pi _1\left( S^1\right) $.\cite
{BP75,Bo78,M79,R82,DEGM82,GM86,S89} Given that $S^1={\bf R}/{\bf Z}$ (${\bf R%
}$ is the additive group of the real numbers), we can write 
\begin{equation}
\label{4.25}\pi _1\left( S^1\right) =\pi _1\left( {\bf R}/{\bf Z}\right)
=\pi _0\left( {\bf Z}\right) ={\bf Z}, 
\end{equation}
which should be compared with (\ref{4.18}). As in the case of layered
superconductors, the external magnetic field $H>0$ breaks the symmetry $%
\varphi \rightarrow -\varphi $. Thus, only the states parameterized by $%
N_v=0,1,\ldots $ are possible, with $N_v=0$ being the ''vacuum'', Meissner
state.

To conclude this discussion, we have to clarify a typical misunderstanding%
\cite{Kr02} concerning the role of the laminar model\cite{dG} in type-II
superconductivity. In reality, the order-parameter space of a continuum
type-II superconductor, $M=S^1$, precludes the existence of topologically
stable plane defects, envisaged by the laminar model. Indeed, consider two
points $P_1=(x_0,a,z_0)$, $P_2=(x_0,b,z_0)$ on the opposite sides of such a
defect, in unperturbed regions of the superconductor. Join these points by a
continuous path, parameterized by $a\leq y\leq b$. The boundary of the
interval $\left[ a,b\right] $ is a $0$-dimensional sphere $S^0=\left\{
a,b\right\} $, which leads us to a consideration of the maps $S^0\rightarrow
S^1$. However, the pertinent homotopy group $\pi _0\left( S^1\right) $ is
trivial,\cite{BP75,Bo78,M79,R82,GM86,S89} i.e., 
\begin{equation}
\label{4.26}\pi _0\left( S^1\right) =0, 
\end{equation}
in contrast to (\ref{4.18}) and (\ref{4.25}). Topological instability of the
''laminar solution'' hardly allows one to expect that this solution
corresponds to any local minimum of the Ginzburg-Landau free-energy
functional. Therefore, a comparison with unstable ''isolated fluxons'' or
''triangular Josephson-vortex lattices'' in layered superconductors is much
more appropriate than the far-fetched\cite{Kr02} ''similarity'' to the
vortex plane.

\subsection{The interpretation of experimental data}

Experimental observations of the vortex structure in layered superconductors
can be roughly subdivided into two groups: (i) direct observations, allowing
one to''visualize'' the flux distribution;\cite{N96,Yu98,Mo98} (ii) indirect
observations (i.e., measurements of $c$-axis transport properties,\cite
{So96,La96,Kr00} magnetization,\cite{De99} and the upper critical field\cite
{Ru80}). Here, we present an overview of these observations, showing that
all the experimental data available up to now can be explained in terms of
the stable vortex-plane configurations. A detailed quantitative analysis can
be done with the use of the results of Refs. \onlinecite{K99,K01,K1}.

Josephson-flux distribution, characteristic of vortex planes for $H>0$, was
directly observed on artificial low-$T_c$ stacked junctions in Ref. \onlinecite{N96} (by low-temperature scanning electron microscopy) and in Ref. \onlinecite{Yu98} (by polarized neutron reflection). In particular, the double-junction
stack\cite{N96} ($N=3$) has revealed the phase-difference symmetry $\phi
_1=\phi _2$, exactly as could be expected from the general relations (\ref
{2.5}) for vortex planes. In both the experiments,\cite{N96,Yu98}
penetration of the flux occurred simultaneously and coherently into all the
junctions, in full agreement with the scenario for the vortex planes.\cite
{K99,K01} Moreover, accompanying measurements of magnetization in Ref. \onlinecite{Yu98} have shown typical oscillations and hysteresis. The oscillations
should be viewed as a manifestation of a series of first-order phase
transitions, discussed in section IV.2 and Refs. \onlinecite{K99,K01}, whereas the
hysteresis is implied by the overlap of the regions (\ref{4.12}), (\ref{4.13}%
).

We draw attention to a possible application to high-$T_c$ superconductivity.
Oscillations of magnetization in parallel fields, interpreted as evidence of
Josephson nature of the flux, have been reported for YBCO in Ref. \onlinecite{De99}%
. According to Ref. \onlinecite{De99}, ''the temperature dependence of the
magnetization contradicts the present theoretical expectations''.

As shown by our self-consistent calculations,\cite{K99,K01} the oscillating
behavior of the critical Josephson current $I_c\left( H\right) $ (the
Fraunhofer pattern) is a result of successive penetration of vortex planes
and their pinning by the edges of the superconductor. Oscillating $I_c\left(
H\right) $ dependencies have been observed both on artificial low-$T_c$
stacked junctions\cite{So96,Kr00} and high-$T_c$ layered superconductors
BSCCO.\cite{La96,Kr00} ''Irregularities'' of the dependence $I_c\left(
H\right) $,\cite{So96,Kr00} such as, e.g., multivaluedness and aperiodicity,
can be easily explained by the overlap of the regions (\ref{4.12}), (\ref
{4.13}). Behavior of this type was observed a long time ago on the single
Josephson junction,\cite{Sch70} which confirmed the theoretical prediction
of the overlap for ordinary Josephson vortices.\cite{OS67}

The most ''ancient'' experimental confirmation of the stability of vortex
planes is provided by observations of the ''crossover'' behavior of $%
H_{c2}\left( T\right) $ in artificial low-$T_c$ stacked junctions.\cite{Ru80}
For $H\preceq H_{c2}$, the condition $f_n=1$, employed in the derivation of
Eqs. (\ref{2.2}), in no longer valid. However, periodic modulations of $%
f_n\left( y\right) $, caused by the presence of vortex planes and therefore
identical in all the S-layers, account for the observed behavior of $%
H_{c2}\left( T\right) $.\cite{K99,K01}

Finally, we want to comment on direct observations of non-equilibrium
isolated vortices in layered high-$T_c$ superconductors at $H=0$.\cite{Mo98}
As is explained at the end of subsection IV.A, nontrivial flux
configurations cannot exist in a layered superconductor with ideal
periodicity at $H=0$, $L<\infty $. However, the presence of structural
defects (e.g., stacking faults, as is hinted in Ref. \onlinecite{Mo98}) violates
the condition of ideal periodicity and should stabilize energetically an
otherwise unstable configuration. In this situation, we indeed expect to
obtain non-equilibrium isolated vortices, because their self-energy is lower
and the $c$-axis extent is smaller than those of vortex planes at $H=0$, $%
L<\infty $.\cite{K01} A detailed mathematical analysis of this case can be
done on the basis of the results of Ref. \onlinecite{K1}.

\section{Conclusions}

In brief, we have solved the problem of the classification of all solutions
to (\ref{2.2}), (\ref{2.9}) with respect to their stability. In our
consideration, we have employed exact methods of the calculus of variations,
soliton physics and the exact results of Ref. \onlinecite{K1} [the expression for
the Gibbs free-energy functional (\ref{3.1}), the first integral (\ref{2.10}%
) and the solution to the soliton boundary value problem (\ref{2.2}), (\ref
{2.4}), (\ref{4.10})]. In view of obvious mathematical complexity, the
problem of the stability of vortex configurations could not be solved by use
of inadequate methods, employed, e.g., in Refs. \onlinecite {B73,CC90,BC91,BLK92,Ko93,Kr02}, and was not even posed in any of these
publications.

In full agreement with the fundamentals of soliton physics, we have proved
that the only minimizers of both the Gibbs and Helmholtz free energy
functionals are the Meissner solution (the ''vacuum'' state) and soliton
vortex-plane solutions. They represent the actual equilibrium field
configurations. The obtained results allowed us to explain exceptional
stability of vortex planes, established in numerical simulations, and to
provide a unified interpretation of the experimental data available up to
now.

In contrast, non-soliton configurations (''isolated fluxons'', triangular
Josephson-vortex lattices'', etc.), proposed in previous publications,
turned out to be absolutely unstable and unobservable: They are nothing but
saddle points of the Gibbs free-energy functional and are not even
stationary points of the Helmholtz free-energy functional. Physically, these
configurations violate conservation laws for the flux and the current.

One may ask a natural question why exactly the unstable configurations were
previously proposed as the ''equilibrium state'', whereas the actual
minimizers of the free energy (vortex-plane solitons) were neglected. We
think that the answer lies in the following:

i) the hypothesis of an ''analogy'' with the Abrikosov-vortex structure in
type-II superconductors, accepted without any mathematical justification;%
\cite{B73}

ii) the absence of an exact mathematical definition of the ''Josephson
vortex''. In Refs. \onlinecite{B73,CC90}, ''isolated Josephson vortices'' were
discussed without any consideration of the SG equations. The fact that
Josephson vortices are nothing but static solitons of the SG equations was
not realized in subsequent publications, either. For example, the existence
of Josephson vortices in the case $W\ll 2\lambda _J$ was denied in Refs. \onlinecite{BCG91,Kr01}, which is refuted by our Eqs. (\ref{4.13}), valid for any $%
W$;

iii) obvious mathematical mistakes in the treatment of the Lawrence-Doniach
model\cite{LD} for infinite ($N=\infty $) layered superconductors. The
neglect\cite{BC91,BCG91,BLK92,Ko93} of the surface variation in the
variational principle for the Lawrence-Doniach functional resulted in a loss
of the conservation laws for the current and the flux, as was first pointed
out in Ref. \onlinecite{K99,K01}: Soliton solutions are a corollary of these
conservation laws;

iv) the absence of any investigation of analytical properties of the coupled
static SG equations for $H>0$, $W<\infty $. Pertinent soliton solutions were
obtained in our papers: in the exactly solvable cases $N=\infty $ (Refs. \onlinecite{K99,K01}), $N=2,3$ (Refs. \onlinecite{K1,K2}), and in the general case $%
2\leq N<\infty $ (Ref. \onlinecite{K1}). Standard methods of soliton physics\cite
{ZK74,BP75,Bo78,R82,DEGM82,GM86,S89} as well as advanced methods of the
calculus of variations and of the theory of differential equations, employed
in our analysis, were completely disregarded in previous theoretical
publications;\cite{BC91,BCG91,BLK92,Ko93}

v) the absence of any attempts to analyze the stability of the proposed
''vortex'' configurations, both in theoretical publications\cite
{B73,CC90,BC91,BLK92,Ko93} and numerical simulations.\cite{Kr01} Dynamic
stability of vortex planes, noticed in numerical simulations,\cite
{SBP93,Kl94,Ma00} was not understood and neglected;

vi) the neglect of direct experimental observations\cite{N96,Yu98} of the
Josephson-vortex structure at $H>0$: These observations have clearly
revealed that exactly the vortex planes (not ''isolated fluxons'' or
''triangular lattices'') are the actual equilibrium field configurations;

vii) long-term domination of the subjective point of view\cite{B73} and the
absence of any pluralism of opinion. As a result, the critical remarks\cite
{F98} are neglected, whereas the attempts to clarify the situation within
the framework of a rigorous mathematical approach\cite{K99,K01} are
immediately attacked\cite{Kr02} with the use of inappropriate methods.\cite
{r1}

We hope that this paper will finally convince both theorists and
experimentalists, specializing in the field of weak superconductivity, of
the necessity to give up the old, unsound theoretical prejudices: The wealth
of magnetic properties of layered superconductors (both low- and high-$T_c$)
cannot be understood without the solitons. One should also think of possible
practical applications of the vortex-plane solitons, e.g., in
submillimeter-wave generators, as is proposed in Ref. \onlinecite{Ma00}. Given the
role of the {\it single} SG equation in different fields of physics (quantum
optics;\cite{DEGM82} the Skyrme and the Thirring models in elementary
particle physics;\cite{Bo78,R82,DEGM82} the theory of dislocations and
magnetism, let alone the Josephson effect, in condensed matter physics,\cite
{DEGM82} etc.), we expect that our exact results for the {\it coupled} SG
equations (including the {\it single} one as a particular case) may find
applications in these fields as well.

The coupled SG equations for $H>0$, $W<\infty $ have not been studied in
mathematical literature, either. Our exact analytical results for the {\it %
static} case constitute only the first step in this direction. The next
stage should be analytical properties of {\it time-dependent} equations. Our
paper may stimulate interest in this problem of specialists in applied
mathematics as well.

\appendix

\section{The properties of the matrix $G\left( n,m\right) $}

The explicit form of $G(n,m)$ is 
\begin{equation}
\label{a1}G(n,m)=\frac 1{2\epsilon \sqrt{1+\frac{\epsilon ^2}4}}\left[ \mu
^{\left| n-m\right| }-\frac{\mu ^n\left( \mu ^{m-N}-\mu ^{N-m}\right) +\mu
^{N-n}\left( \mu ^{-m}-\mu ^m\right) }{\mu ^{-N}-\mu ^N}\right] , 
\end{equation}
where 
\begin{equation}
\label{a2}\mu =1+\frac{\epsilon ^2}2-\epsilon \sqrt{1+\frac{\epsilon ^2}4}. 
\end{equation}
The following properties of $G(n,m)$ are obvious: 
\begin{equation}
\label{a3}G(n,m)=G(m,n), 
\end{equation}
\begin{equation}
\label{a4}G(n,N-m)=G(N-n,m). 
\end{equation}
The matrix $G(n,m)$ is positive definite, since all its eigenvalues $e_j$
are positive: 
\begin{equation}
\label{a5}e_j=\frac{\lambda _j^2}{\epsilon ^2},\qquad \lambda _j=\frac
\epsilon {\sqrt{2+\epsilon ^2-2\cos \frac{\pi j}N}},\qquad j=1,2,\ldots
,N-1. 
\end{equation}
Of importance are the summation rules: 
$$
\sum_{m=1}^{N-1}G(n,m)=\frac 1{\epsilon ^2}\left[ 1-G(n,1)-G(n,N-1)\right] 
$$
\begin{equation}
\label{a5.1}=\frac 1{\epsilon ^2}\left[ 1-\frac{\mu ^{-n}+\mu ^{-N+n}-\mu
^n-\mu ^{N-n}}{\mu ^{-N}-\mu ^N}\right] ,\quad 1\leq n\leq N-1; 
\end{equation}
\begin{equation}
\label{a6}\sum_{n=1}^{N-1}\sum_{m=1}^{N-1}G(n,m)=\frac 1{\epsilon ^2}\left[
N-1-\frac{2\sqrt{1+\frac{\epsilon ^2}4}-\epsilon }\epsilon \frac{1-\mu
^{N-1} }{1+\mu ^N}\right] \equiv \frac{N-1}{\epsilon ^2H_s^2}, 
\end{equation}
where%
$$
H_s=\left[ 1-\frac{2\sqrt{1+\frac{\epsilon ^2}4}-\epsilon }{\epsilon \left(
N-1\right) }\frac{1-\mu ^{N-1}}{1+\mu ^N}\right] ^{-\frac 12} 
$$
\begin{equation}
\label{a7}\equiv \sqrt{\frac{\left( N-1\right) N}2}\left[ \sum_{k=0}^{\left[
\frac N2\right] -1}\lambda _{2k+1}^2\cos {}^2\frac{\pi \left( 2k+1\right) }{%
2N}\right] ^{-\frac 12} 
\end{equation}
is the superheating (penetration) field of a semiinfinite ($0\leq y<\infty $%
) layered superconductor.\cite{K1}

\section{A relationship to the variational principle for infinite layered
superconductors ($N=\infty $)}

The intralayer currents for $f_n=1$ are given by\cite{K1}%
$$
J_n\left( y\right) =\frac{d\varphi _n\left( y\right) }{dy}-2A\left(
n,y\right) 
$$
\begin{equation}
\label{b1}=\frac 1{4\pi }\left[ H_n\left( y\right) -H_{n+1}\left( y\right)
\right] ,\quad n=0,1,\ldots ,N-1, 
\end{equation}
where $H_0\left( y\right) =H_N\left( y\right) \equiv H$. Using the second
relation in (\ref{b1}), we get 
\begin{equation}
\label{b2}\sum_{n=0}^{N-1}J_n\left( y\right) =0, 
\end{equation}
which is the conservation law for the total intralayer current. Moreover, in
view of (\ref{2.5}) and (\ref{a3}), (\ref{a4}), 
\begin{equation}
\label{b3}H_n\left( y\right) =H_{N-n}\left( y\right) . 
\end{equation}
Hence, 
\begin{equation}
\label{b4}J_n\left( y\right) =-J_{N-n-1}\left( y\right) . 
\end{equation}

Using the first relation in (\ref{b1}), we write%
$$
\int\limits_{-L}^Ldy\left[ J_n\left( y\right) -J_{n-1}\left( y\right)
\right] 
$$
\begin{equation}
\label{b5}=\phi _n\left( L\right) -\phi _n\left( -L\right)
-\int\limits_{-L}^Ldy\left[ A\left( n,y\right) -A\left( n-1,y\right) \right]
,\quad n=1,2,\ldots ,N-1. 
\end{equation}
The second term on the right-hand side of (\ref{b5}) is the flux between the
S-layers $n$ and $n-1$. We can therefore rewrite (\ref{b5}) using (\ref{2.8}%
):%
$$
\int\limits_{-L}^Ldy\left[ J_n\left( y\right) -J_{n-1}\left( y\right)
\right] =HW\left[ G\left( n,1\right) +G\left( n,N-1\right) \right] 
$$
\begin{equation}
\label{b6}+\phi _n\left( L\right) -\phi _n\left( -L\right) +\frac{\epsilon ^2%
}2\sum_{m=1}^{N-1}G\left( n,m\right) \left[ \phi _m\left( L\right) -\phi
_m\left( -L\right) \right] ,\quad n=1,2,\ldots ,N-1. 
\end{equation}
In view of the flux-conservation conditions (\ref{3.13}), the variation of
the right-hand side of (\ref{b6}) vanishes, hence 
\begin{equation}
\label{b7}\delta J_n\left( y\right) =\delta J_{n-1}\left( y\right) ,\quad
n=1,2,\ldots ,N-1. 
\end{equation}
Combined with the current-conservation law (\ref{b2}), relations (\ref{b7})
yield: 
\begin{equation}
\label{b8}\delta J_n\left( y\right) =0,\quad n=0,1,\ldots ,N-1, 
\end{equation}
which means that partial intralayer currents are also conserved.

Consider the case $N\gg 1$. For $n$ satisfying the condition $\left[
\epsilon ^{-1}\right] \ll n\ll N-1-\left[ \epsilon ^{-1}\right] $, we can
proceed to the limit $N\rightarrow \infty $ in the second relation (\ref{b1}%
), obtaining 
\begin{equation}
\label{b9}J_n\left( y\right) =\frac{d\varphi _n\left( y\right) }{dy}%
-2A\left( n,y\right) =0. 
\end{equation}
This is exactly the result derived for the infinite ($N=\infty $) layered
superconductor in Refs. \onlinecite{K99,K01} by means of an exact variational
principle, based on the use of the conservation law for the total intralayer
current.


\begin{thebibliography}{99}
\bibitem{K99}  S. V. Kuplevakhsky, Phys. Rev. B {\bf 60}, 7496 (1999).

\bibitem{K01}  S. V. Kuplevakhsky, Phys. Rev. B {\bf 63}, 054508 (2001).

\bibitem{K1}  S. V. Kuplevakhsky, cond-mat/0202293 (submitted to Phys. Rev.
B).

\bibitem{BP75}  A. A. Belavin and A. M. Polyakov, Pis'ma Zh. Eksp. Teor.
Fiz. {\bf 22}, 245 (1975).

\bibitem{Bo78}  L. J. Boya, J. F. Carine\~na, and J. Mateos, Fortschritte
der Physik, {\bf 26}, 175 (1978).

\bibitem{R82}  R. Rajaraman, {\it Solitons and Instantons} (North-Holland,
Amsterdam, 1982).

\bibitem{DEGM82}  R. K. Dodd, J. C. Eilbeck, J. D. Gibbon, and H. C. Morris, 
{\it Solitons and Nonlinear Wave Equations} (Academic Press, London, 1982).

\bibitem{GM86}  P. Goddard and P. Mansfield, Rep. Prog. Phys. {\bf 49}, 725
(1986).

\bibitem{S89}  A. S. Schwarz, {\it Quantum Field Theory and Topology}
(Nauka, Moscow, 1989) (in Russian).

\bibitem{r2}  The conserved topological charge of soliton solutions (the
flux in our case) provides lower-bound estimates for the energy functionals.

\bibitem{B73}  L. N. Bulaevskii, Zh. Eksp. Teor. Fiz. {\bf 64}, 2241 (1973)
[Sov. Phys. JETP {\bf 37}, 1133 (1973)].

\bibitem{Kr02}  V. M. Krasnov, Phys. Rev. B {\bf 65}, 096503 (2002).

\bibitem{r1}  We cannot but mention a deplorable fact that the exact
mathematical results of Refs. \onlinecite{K99,K01} and the actual methods of their
derivation are grossly misrepresented in Ref. \onlinecite{Kr02}: see our reply,
Ref. \onlinecite{K2}.

\bibitem{dG}  P. G. DeGennes, {\it Superconductivity of Metals and Alloys}
(Benjamin, New York, 1966).

\bibitem{BCG91}  L. N. Bulaevskii, J. R. Clem, and L. I. Glazman, Phys. Rev.
B {\bf 46}, 350 (1991).

\bibitem{Kr01}  V. M. Krasnov, Phys. Rev. B {\bf 63}, 064519 (2001).

\bibitem{K2}  S. V. Kuplevakhsky, cond-mat/0204170 (submitted to Phys. Rev.
B).

\bibitem{CC90}  J. R. Clem and M. W. Coffey, Phys. Rev. B {\bf 42}, 6209
(1990).

\bibitem{BC91}  L. N. Bulaevskii and J. R. Clem, Phys. Rev. B {\bf 44},
10234 (1991).

\bibitem{BLK92}  L. N. Bulaevskii, M. Ledvij and V. G. Kogan, Phys. Rev. B 
{\bf 46}, 366 (1992).

\bibitem{Ko93}  A. E. Koshelev, Phys. Rev. B {\bf 48}, 1180 (1993).

\bibitem{F98}  B. Farid, J. Phys. Condens. Matter {\bf 10}, L589 (1998).

\bibitem{N96}  I. P. Nevirkovets, T. Doderer, A. Laub, M. G. Blamire, and J.
E. Evetts, J. Appl. Phys. {\bf 80}, 2321 (1996).

\bibitem{Yu98}  S. M. Yusuf, E. E. Fullerton, R. M. Osgood II, and G. P.
Felcher, J. Appl. Phys. {\bf 83}, 6801 (1998); S. M. Yusuf, R. M. Osgood
III, J. S. Jiang, C. H. Sowers, S. D. Bader, E. E. Fullerton, and G. P.
Felcher, J. Magn. Magn. Mater. {\bf 198}-{\bf 199}, 564 (1999).

\bibitem{Kr00}  V. M. Krasnov, V. A. Oboznov, V. V. Ryazanov, N. Mros, A.
Yurgens, and D. Winkler Phys. Rev. B {\bf 61}, 766 (2000).

\bibitem{CH}  According to, e.g., R. Curant and D. Hilbert, {\it Methods of
Mathematical Physics}, Vol. II (Interscience, New York, 1962), the {\it %
uniqueness} of the solution is one of the three criteria of the correctness
of the formulation of the boundary value problem (the other two are the {\it %
existence} of the solution and its {\it continuous dependence} on the
boundary conditions).

\bibitem{L62}  C. Lancsos, {\it The Variational Principles of Mechanics}
(University of Toronto Press, Toronto, 1962).

\bibitem{ZK74}  V. E. Zakharov and E. A. Kuznetsov, Zh. Eksp. Teor. Fiz. 
{\bf 66}, 594 (1974).

\bibitem{M79}  N. D. Mermin, Rev. Mod. Phys., {\bf 51}, 591 (1979).

\bibitem{SBP93}  S. Sakai, P. Bodin, and N. F. Pedersen, J. Appl. Phys. {\bf %
73}, 2411 (1993).

\bibitem{Kl94}  R. Kleiner, P. M\"uller, H. Kohlstedt, N. F. Pedersen, and
S. Sakai, Phys. Rev. B {\bf 50}, 3942 (1994).

\bibitem{Ma00}  M. Machida, T. Koyama, A. Tanaka, and M. Tachiki, Physica C 
{\bf 330}, 85 (2000).

\bibitem{LL}  L. D. Landau and E. M. Lifshitz, {\it Electrodynamics of the
Continuum Media}, (Pergamon, Oxford, 1983).

\bibitem{OS67}  C. S. Owen and D. J. Scalapino, Phys. Rev. {\bf 164}, 538
(1967).

\bibitem{K70}  I. O. Kulik and I. K. Yanson, {\it The Josephson Effect in
Superconducting Tunneling Structures} (Israel Program for Scientific
Translation, Jerusalem, 1972).

\bibitem{BF92}  A. Buzdin and D. Feinberg, Phys. Lett. A {\bf 165}, 281
(1992).

\bibitem{Mo98}  K. A. Moler, J. R. Kirtley, D. G. Hinks, T. W. Li, and M.
Xu, Science {\bf 279}, 1193 (1998).

\bibitem{So96}  S. N. Song, P. R. Auvil, M. Ulmer, and J. B. Ketterson,
Phys. Rev. B {\bf 53}, R6018 (1996).

\bibitem{La96}  Yu. I. Latyshev, J. E. Nevelskaya, and P. Monceau, Phys.
Rev. Lett. {\bf 77}, 932 (1996).

\bibitem{De99}  K. Deligiannis, S. Kokkaliaris, M. Oussena, P. A. J. de
Groot, L. Fr\"uchter, R. Gagnon, and L Taillefer, Phys. Rev. B {\bf 59},
14772 (1999).

\bibitem{Ru80}  S. T. Ruggiero, T. W. Barbee, Jr., and M. R. Beasly, Phys.
Rev. Lett. {\bf 45}, 1299 (1980).

\bibitem{Sch70}  K. Schwidtal, Phys. Rev. B {\bf 23}, 2526 (1970).

\bibitem{LD}  W. E. Lawrence and S. Doniach, in {\it Proceedings of the
Twelfth Conference on Low Temperature Physics, Kyoto, 1970, }edited by E.
Kanda (Keigaku, Tokyo, 1970), p. 361.
\end{thebibliography}
\end{document}